\documentclass[12pt]{article}

\makeatletter
 
\@addtoreset{equation}{section}
\makeatother

\newcommand{\q}{\quad}
\newcommand{\ts}{\hspace*}

\newcommand{\hs}{\hspace}
\newcommand{\vs}{\vspace}
\newcommand{\bc}{\begin{center}}
\newcommand{\ec}{\end{center}}
\newcommand{\f}{\frac}
\newcommand{\dps}{\displaystyle}

\newcommand{\Phat}{\hat{{\cal P}}}
\newcommand{\Qhat}{\hat{\Omega}}
\newcommand{\Zp}{\hat{Z}^{(+)}}
\newcommand{\Zm}{\hat{Z}^{(-)}}
\newcommand{\Zpm}{\hat{Z}^{(\pm)}}
\newcommand{\Zmp}{\hat{Z}^{(\mp)}}
\newcommand{\Ap}{\hat{A}^{(+)}}
\newcommand{\Am}{\hat{A}^{(-)}}
\newcommand{\Apm}{\hat{A}^{(\pm)}}

\newcommand{\Bp}{\hat{B}^{(+)}}
\newcommand{\Bm}{\hat{B}^{(-)}}

\newcommand{\Bmp}{\hat{B}^{(\mp)}}

\newcommand{\calC}{{\cal C}}

\newcommand{\calO}{{\cal O}}

\newcommand{\hope}{{\it hyper}-operator}

\newcommand{\pscrp}{\mbox{{\scriptsize $\Phat$}}}

\newcommand{\partd}[2]{\f{\partial #1}{\partial #2}}
\newcommand{\commu}[2]{[#1,#2]}

\newcommand{\symmp}[2]{ \{#1,#2\}}

\begin{document}

\begin{center}
{\bf \Large Star-product Description of Quantization in Second-class Constraint Systems}\vs{12pt}\\
M. Nakamura\footnote{\label{*}E-mail:mnakamur@hm.tokoha-u.ac.jp}
\vs{12pt}\\
{\it Research Institute, Hamamatsu Campus, Tokoha University, Miyakoda-cho 1230, 
Kita-ku, Hamamastu-shi, Shizuoka 431-2102, Japan}
\end{center}

\begin{abstract}
The quantization of the second-class constraint systems is  discussed within the projection operator method(POM) of constraint systems. Through the nonlocal representation of the constraint {\it hyper}-operators, new star-products are defined. Then, the projected operator-algebra of the quantized constraint systems is constructed with these star-products, and it is shown that the commutators and symmetrized products among the projected operators contain  the quantum corrections due to the noncommutativity among operators. \vs{12pt} 
\end{abstract} 

\section{Introduction}

\ts{12pt} The problem of the quantization of constraint systems has been extensively investigated as one of the most fundamental problems in gauge theories \cite{HT,HRT}. There exist two standard approaches to quantize constraint systems. The first approach, which we shall call the approach I, is to impose the constraints first and then to quantize on the reduced phase space \cite{DC}. The second, which we shall call the approach II, is inversely first to quantize on the unconstrained phase space, and then to impose the constraints as the operator-equations \cite{BF1,BF2}.\\
\ts{12pt}There often occur the situations where the two are not equivalent. In the case of the first-class constraints, this problem has been extensively investigated until now , and it has been shown that the approach II involves the contributions which can never appear in the approach I \cite{Loll,DJT,Witten,Jor}.\\
\ts{12pt}In the case of second-class constraints, the quantization of the constraint systems in the canonical formalism has usually been accomplished by using the generalized Hamiltonian formalism with the Dirac bracket \cite{DC}, which is in the approach I. There have been proposed many 
formalism within approach II; the operator formalism  developed by 
Batalin and Fradkin \cite{BF1,BF2,BF3}, the projection operator formalism \cite{MN1,MN2,MN3,MN4,BLM,KRF}, the algebraic approach proposed by Ohnuki and Kitakado \cite{OK} and the first-order singular Lagrangian formalism proposed by Faddeev and Jackiw \cite{FJ,J}. Using the projection operator method \cite{MN1,MN2} (POM), then, we have shown that the operator-algebra in the approach II contain the quantum corrections caused by the noncommutativity in the re-ordering of constraint-operators in the products of operators, which can never be obatined by the approach I.  Alternative formalism of quantization is the Wely-Wigner-Moyal (WWM) quantization \cite{WW,Moy}, and has been investigated for the constraint systems \cite{BF4,HKM,KOS,Kr}. In the WWM, the star-product is defined in the following form:
$$
\star=\exp\left(\f{i\hbar}2\Lambda^{\mu\nu}\overleftarrow{\partial}_{\mu}\overrightarrow{\partial}_{\nu}\right),
\eqno{(1.1)}
$$
where $\Lambda^{\mu\nu}$ is the appropriative symplectic matrix \cite{Moy,BF4,Gr,Bastos}.\\
\ts{12pt}By using the nonlocal representation of the operations of {\it hyper}-operators\footnote{We shall call operators acting on a vector space of linear operarors {\it hyper}-operators.} $\hat{\calO}$ and $\hat{\calO}'$ on the operators $X$ and $Y$ with the operation-indices $k$,$l$,
$$
(\hat{\calO}X)(\hat{\calO}'Y)=\hat{\calO}(k)\hat{\calO}'(l)X(k)Y(l)|_{k=l},
\eqno{(1.2)}
$$
we shall propose new star-products, which are composed with the constraint {\it hyper}-operators and have the same structures as (1.1). Then, the projected operator-algebra formulated in the POM is reformulated in terms of these new star-products and the reformulated algebra is shown to be the extension of the Moyal star-products to the operator formalism in the constraint systems.\\
\ts{12pt}The present paper is organized as follows. In sect. 2, we review the POM in terms of the symplectic representation.
In sect. 3, we introduce the nonlocal representation for the operations of {\it hyper}-operators in the form of (1.2) and new star-products are proposed. By using these star-products, then, we present the general formulas of the commutators and  symmetrized products among the projected operators in the second-class constraint system. It is shown that these commutators and symmetrized products contain the quantum correction terms due to the noncommutativity of  operators. In sect. 4, some concluding remarks are given.
   
%
%
\section{Projection Operator Method of Constraint System}

\ts{12pt}We shall treat the constraint systems with the supersymmetry. So, we will adopt the graded commutator defined as
$$
\commu{A}{B}=AB-(-1)^{\epsilon(A)\epsilon(B)} BA
\eqno{(2.1)}
$$
and the graded symmetrized product 
$$
\symmp{A}{B}=\f12(AB+(-1)^{\epsilon(A)\epsilon(B)}BA)
\eqno{(2.2)}
$$
for operators $A$ and $B$, where $\epsilon(A)$ is the Grassmann parity of $A$. Let $\Ap$, $\Am$ be the {\it hyper}-operators defined, respectively, as follows: For any operator $O$,
 $$
 \Ap O=\symmp{A}{O}, \hs{24pt}\Am O=\f1{i\hbar}\commu{A}{O}.
 \eqno{(2.3)}
 $$
Then, they obey the {\it hyper}-commutator algebra
$$
\begin{array}{l}
\commu{\Ap}{\Bp}=\dps{\f{i\hbar}4}\hat{C}^{(-)}_{AB},\vs{12pt}\\
\commu{\Am}{\Bm}=\dps{\f1{i\hbar}}\hat{C}^{(-)}_{AB},\vs{12pt}\\
\commu{\Apm}{\Bmp}=\dps{\f1{i\hbar}}\hat{C}^{(+)}_{AB},
\end{array}
\eqno{(2.4)}
$$ 
where $C_{AB}=\commu{A}{B}$.

\subsection{Second-class constraint system}

\ts{12pt}Let $(\calC,H(\calC),T_{\alpha}(\calC))$ be the initial unconstraint quantum system, where $\calC=\{(q^i,p_i);i=1,\cdots ,N\}$ is a set of canonically conjugate operators (CCS) satisfying the canonical commutation relations (CCR)
$$
\commu{q^i}{p_j}=i\hbar\delta^i_j,\hs{36pt}\commu{q^i}{q^j}=\commu{p_i}{p_j}=0,
\eqno{(2.5)}
$$ 
$H(\calC)$, the Hamiltonian of the initial unconstraint system and $T_{\alpha}(\calC)$ $(\alpha=1,\cdots ,2M<2N)$, the constraint-operators corresponding to the second-class constraints $T_{\alpha}=0$. Starting with $(\calC,H(\calC),T_{\alpha}(\calC))$, then, we construct the constraint quantum system  $(\calC^*,H^*(\calC^*),T_{\alpha}(\calC^*))$, where  $\calC^*$ is the set of $N-M$ canonically conjugate pairs stisfying
$$
T_{\alpha}(\calC^*)=0\hs{48pt}(\alpha=1,\cdots,2M).
\eqno{(2.6)}
$$
\ts{12pt}From the Darboux's theorem in dynamical systems, we can, in principle, construct the canonically conjugate set in terms of $T_{\alpha}(\calC)$, which we call the associated canonically conjugate set (ACCS) \cite{MN1,MN2,MN3}. \\
\ts{12pt}Let $\{(\xi^a,\pi_a)|\epsilon(\xi^a)=\epsilon(\pi_a)=s,a=1,\cdots,M\}$ be the ACCS constructed with $T_{\alpha}$. In order to collectively represent the ACCS, we introduce the symplectic form $Z_{\alpha}$ as follows:
$$ 
Z_{\alpha}=\left\{\begin{array}{l}\xi^a\hs{24pt}(\alpha=a)\vs{6pt}\\
\pi_a\hs{24pt}(\alpha=a+M) \hs{36pt}(\alpha=1,\cdots,2M\q;\q a=1,\cdots,M),
\end{array}\right.
\eqno{(2.7)}
$$
which obeys the commutation relation
$$
[Z_{\alpha},\ Z_{\beta}]=i\hbar(- (-)^s)J_{\alpha\beta}=i\hbar J^{\alpha\beta},
\eqno{(2.8)}
$$
where
$$
J^{\alpha\beta}=\left(\begin{array}{rr}O&I\vs{6pt}\\
-(-)^sI&O\end{array}\right)_{\alpha\beta}\hs{24pt}\mbox{with}\left\{\begin{array}{l}I:M\times M\ \mbox{unit matrix}\vs{6pt}\\
       O:M\times M\ \mbox{zero matrix}\end{array}\right.
\eqno{(2.9a)}
$$
is the supersymmetric symplectic matrix and $J_{\alpha\beta}$ 
is the inverse of $J^{\alpha\beta}$, which satisfies 
$$  J_{\alpha\beta}=-(-1)^sJ^{\alpha\beta}=J^{\beta\alpha}.
\eqno{(2.9b)}
$$
From (2.4) and (2.8), the symplectic \hope s $\Zpm$ obey the {\it hyper}-commutation relations
$$
\begin{array}{l}
[\Zpm_{\alpha},\Zpm_{\beta}]=0,\vs{6pt}\\
\commu{\Zpm_{\alpha}}{\Zmp_{\beta}}=\commu{\Zmp_{\alpha}}{\Zpm_{\beta}}=J^{\alpha\beta}.
\end{array}
\eqno{(2.10)}
$$ 
\subsection{Projection operator $\Phat$  and projected system}

\ts{12pt}Let $\Phat$ be the \hope\ defined by
$$
\begin{array}{rcl}
\Phat&=&\dps{\sum^{\infty}_{n=0}\f{(-)^{ns}}{n!}J^{\alpha_1\beta_1}\cdots J^{\alpha_n\beta_n}\Zp_{\alpha_1}\cdots\Zp_{\alpha_n}\Zm_{\beta_n}\cdots \Zm_{\beta_1}}\vs{12pt}\\
&=&\dps{\exp\left[(-1)^s\Zp_{\alpha}\f{\partial}{\partial\varphi_{\alpha}}\right]\exp[J^{\alpha\beta}\varphi_{\alpha}\Zm_{\beta}]|_{\varphi=0}},
\end{array}
\eqno{(2.11)}
$$
which satisfies the following algebraic properties \cite{MN2,MN3,MN4}:\vs{3pt}\\
$$
\Phat\cdot\Phat=\Phat,
\eqno{(2.12)}
$$

$$
\begin{array}{l}
\dps{\sum^{\infty}_{n=0}\f{(-1)^{(s+1)n}}{n!}J^{\alpha_1\beta_1}\cdots J^{\alpha_n\beta_n}\Zp_{\alpha_1}\cdots \Zp_{\alpha_n}\Phat \Zm_{\beta_n}\cdots \Zm_{\beta_1}}\vs{12pt}\\
=\dps{\exp\left[-(-1)^s\Zp_{\alpha}\partd{}{\varphi_{\alpha}}\right]\Phat\exp[J^{\alpha\beta}\varphi_{\alpha}\Zm_{\beta}]|_{\varphi=0}}\vs{12pt}\\
=\dps{\exp\left[(-1)^s\Zp_{\alpha}\partd{}{\varphi_{\alpha}}\right]\Phat\exp[-J^{\alpha\beta}\varphi_{\alpha}\Zm_{\beta}]|_{\varphi=0}}=1,
\end{array}
\eqno{(2.13)}
$$
\vs{3pt}\\
$$
\Phat Z_{\alpha}=0,\hs{36pt}\Phat\Zp_{\alpha}=\Zm_{\alpha}\Phat=0.
\eqno{(2.14)}
$$\vs{3pt}\\
Then, the available formulas for the projections of commutators and symmetrized products including $\Zp$ are obtained, which we shall present in Appendix A. \\
\ts{12pt}Through the operation of $\Phat$, now,  the initial unconstraint system $(\calC,H(\calC),T_{\alpha}(\calC))$ is projected on to the constraint system $(\calC^*,H^*(\calC^*),T_{\alpha}(\calC^*))$ as follows:
$$
(\calC,H(\calC),T_{\alpha}(\calC))\longmapsto (\calC^*,H^*(\calC^*),T_{\alpha}(\calC^*)),
\eqno{(2.15)}
$$
and the operator $O\in \calC$ is transformed to 
$O^*=\Phat O\in \calC^*$, where
$$
\begin{array}{l}
\calC^*=\Phat\calC=\{(q^{*1},p_{1}^*),\cdots,(q^{*N^*},p^*_{N^*})\},\hs{36pt}(N^*=N-M),\vs{6pt}\\
H^*(\calC^*)=\Phat H(\calC)=H^*(\Phat\calC),\vs{6pt}\\
T_{\alpha}(\calC^*)=\Phat T_{\alpha}(\calC)=0.
\end{array}
\eqno{(2.16)}
$$
The conditions $T_{\alpha}(\calC^*)=0$ $(\alpha=1,\cdots,2M)$ are called as the projection conditions \cite{MN3}.

\subsection{Commutator and Symmetrized product formulas of projected operators}
\ts{12pt}By using the algebraic properties (2.12)-(2.14) and (A.1)-(A.4), the following formulas of the commutators and the symmetrized products in the projected constraint system $(\calC^*,H^*(\calC^*),T_{\alpha}(\calC^*))$ have been obtained \cite{MN2,MN3,MN4}:
$$
\begin{array}{l}
\commu{\Phat X}{\Phat Y}=\dps{\sum^{\infty}_{n=0}(-)^{ns}C_{2n}J^{\alpha_1\beta_1}\cdots J^{\alpha_{2n}\beta_{2n}}\Phat\commu{\Zm_{\alpha_{2n}}\cdots\Zm_{\alpha_1}X}{\Zm_{\beta_{2n}}\cdots\Zm_{\beta_1}Y}}\vs{12pt}\\
+\dps{2(-)^{\varepsilon_Xs+s}\sum^{\infty}_{n=0}(-)^{ns}C_{2n+1}J^{\alpha_1\beta_1}\cdots J^{\alpha_{2n+1}\beta_{2n+1}}\Phat\symmp{\Zm_{\alpha_{2n+1}}\cdots\Zm_{\alpha_1}X}{\Zm_{\beta_{2n+1}}\cdots\Zm_{\beta_1}Y}},
\end{array}
\vs{12pt}
\eqno{(2.17a)}
$$
$$
\begin{array}{l}
\symmp{\Phat X}{\Phat Y}=\dps{\sum^{\infty}_{n=0}(-)^{ns}C_{2n}J^{\alpha_1\beta_1}\cdots J^{\alpha_{2n}\beta_{2n}}\Phat\symmp{\Zm_{\alpha_{2n}}\cdots\Zm_{\alpha_1}X}{\Zm_{\beta_{2n}}\cdots\Zm_{\beta_1}Y}}\vs{12pt}\\
+\dps{\f12(-)^{\varepsilon_Xs+s}\sum^{\infty}_{n=0}(-)^{ns}C_{2n+1}J^{\alpha_1\beta_1}\cdots J^{\alpha_{2n+1}\beta_{2n+1}}\Phat\commu{\Zm_{\alpha_{2n+1}}\cdots\Zm_{\alpha_1}X}{\Zm_{\beta_{2n+1}}\cdots\Zm_{\beta_1}Y}},
\end{array}
\vs{12pt}
\eqno{(2.17b)}
$$
$$
\begin{array}{l}
\Phat\commu{X}{Y}=\dps{\sum^{\infty}_{n=0}(-)^{ns}C_{2n}J^{\alpha_1\beta_1}\cdots J^{\alpha_{2n}\beta_{2n}}\commu{\Phat\Zm_{\alpha_{2n}}\cdots\Zm_{\alpha_1}X}{\Phat\Zm_{\beta_{2n}}\cdots\Zm_{\beta_1}Y}}\vs{12pt}\\
-\dps{2(-)^{\varepsilon_Xs+s}\sum^{\infty}_{n=0}(-)^{ns}C_{2n+1}J^{\alpha_1\beta_1}\cdots J^{\alpha_{2n+1}\beta_{2n+1}}\symmp{\Phat\Zm_{\alpha_{2n+1}}\cdots\Zm_{\alpha_1}X}{\Phat\Zm_{\beta_{2n+1}}\cdots\Zm_{\beta_1}Y}},
\end{array}
\vs{12pt}
\eqno{(2.18a)}
$$
$$
\begin{array}{l}
\Phat\symmp{X}{Y}=\dps{\sum^{\infty}_{n=0}(-)^{ns}C_{2n}J^{\alpha_1\beta_1}\cdots J^{\alpha_{2n}\beta_{2n}}\symmp{\Phat\Zm_{\alpha_{2n}}\cdots\Zm_{\alpha_1}X}{\Phat\Zm_{\beta_{2n}}\cdots\Zm_{\beta_1}Y}}\vs{12pt}\\
-\dps{\f12(-)^{\varepsilon_Xs+s}\sum^{\infty}_{n=0}(-)^{ns}C_{2n+1}J^{\alpha_1\beta_1}\cdots J^{\alpha_{2n+1}\beta_{2n+1}}\commu{\Phat\Zm_{\alpha_{2n+1}}\cdots\Zm_{\alpha_1}X}{\Phat\Zm_{\beta_{2n+1}}\cdots\Zm_{\beta_1}Y}},
\end{array}
\vs{12pt}
\eqno{(2.18b)}
$$
where
$$
C_{n}=\f1{n!}\left(\f{\hbar}{2i}\right)^{n}.
\eqno{(2.19)}
$$


\section{Star-product Representaion of Projected operator Algebra}


\subsection{Nonlocal representation of Projected operator Algebra}

\ts{12pt}Let the operator $\calO$ in the CCS with the operation-index $\eta$, $\calC_{\eta}$,
 be denoted by $\calO(\eta)$. Then, the products of such operators as  $\Zm_{\alpha_{n}}\cdots\Zm_{\alpha_1}X$ are expressed in the nonlocal representations with straightforwad calculations as follows:
$$
(\Zm_{\alpha_{2n}}\cdots\Zm_{\alpha_1}X)(\Zm_{\beta_{2n}}\cdots\Zm_{\beta_1}Y)
=(-1)^{ns}\left.(\Zm_{\alpha}(\eta)\Zm_{\beta}(\zeta))^{2n}X(\eta)Y(\zeta)\right|_{\eta=\zeta},
\vs{12pt}
\eqno{(3.1a)}
$$
$$
(\Zm_{\alpha_{2n+1}}\cdots\Zm_{\alpha_1}X)(\Zm_{\beta_{2n+1}}\cdots\Zm_{\beta_1}Y)
=(-1)^{ns+\varepsilon_Xs}\left.(\Zm_{\alpha}(\eta)\Zm_{\beta}(\zeta))^{2n+1}X(\eta)Y(\zeta)\right|_{\eta=\zeta},
\vs{12pt}
\eqno{(3.1b)}
$$
where
$$
(\Zm_{\alpha}(\eta)\Zm_{\beta}(\zeta))^n=(\Zm_{\alpha_1}(\eta)\Zm_{\beta_1}(\zeta))\cdots(\Zm_{\alpha_n}(\eta)\Zm_{\beta_n}(\zeta))
\eqno{(3.1c)}
$$
and the product of the projected operators becomes as
$$
(\Phat X)(\Phat Y)=\left.\left(\Phat(\eta)\Phat(\zeta)X(\eta)Y(\zeta)\right)\right|_{\eta=\zeta}.
\eqno{(3.2)}
$$

\subsubsection{Commutator and Symmetrized product of projected operators}

\ts{12pt}Let us now introduce the nonlocal {\it hyper}-operator 
$$
\Qhat_{\eta\zeta}=(-1)^s\Zm_{\alpha}(\eta)J^{\alpha\beta}\Zm_{\beta}(\zeta)
\eqno{(3.3)}
$$
for the ACCS $\{Z_{\alpha}\}$ with $\epsilon(Z_{\alpha})=s$.\\
\ts{12pt}By using the formulas (3.1a), (3.1b), (3.2) and (3.3), then, one obtain the nonlocal representations of the commutator (2.17a) and the symmetrized product (2.17b) in the following way:
$$
\begin{array}{l}

\commu{\Phat X}{\Phat Y}\vs{12pt}\\

=\dps{\Phat\left(\left.\sum^{\infty}_{n=0}C_{2n}(J^{\alpha\beta}\Zm_{\alpha}(\eta)\Zm_{\beta}(\zeta))^{2n}\commu{X(\eta)}{Y(\zeta)}\right|_{\eta=\zeta}\right.}\vs{12pt}\\

\ts{24pt}+\dps{\left.\left.(-1)^s\sum^{\infty}_{n=0}C_{2n+1}(J^{\alpha\beta}\Zm_{\alpha}(\eta)\Zm_{\beta}(\zeta))^{2n+1}2\symmp{X(\eta)}{Y(\zeta)}\right|_{\eta=\zeta}\right)}\vs{12pt}\\

=\dps{\Phat\left(\left.\sum^{\infty}_{n=0}\left(\f1{(2n)!}(\f{\hbar}{2i}\Qhat_{\eta\zeta})^{2n}\commu{X(\eta)}{Y(\zeta)}+\f1{(2n+1)!}(\f{\hbar}{2i}\Qhat_{\eta\zeta})^{2n+1}2\symmp{X(\eta)}{Y(\zeta)}\right)\right|_{\eta=\zeta}\right)},

\end{array}
\eqno{(3.4a)}
$$
and similarly,
$$
\begin{array}{l}

\symmp{\Phat X}{\Phat Y}\vs{12pt}\\

=\dps{\Phat\left(\left.\sum^{\infty}_{n=0}\left(\f1{(2n)!}(\f{\hbar}{2i}\Qhat_{\eta\zeta})^{2n}\symmp{X(\eta)}{Y(\zeta)}+\f1{(2n+1)!}(\f{\hbar}{2i}\Qhat_{\eta\zeta})^{2n+1}\f12\commu{X(\eta)}{Y(\zeta)}\right)\right|_{\eta=\zeta}\right)}.

\end{array}
\eqno{(3.4b)}
$$

\subsubsection{Projections of Commutator and symmetrized product}

\ts{12pt}Let us now denote $\Qhat_{\zeta\eta}$ by the 'transpose' $\Qhat^t_{\eta\zeta}$, which, from (2.9c),   satisfies 
$$
\Qhat^t_{\eta\zeta}=\Qhat_{\zeta\eta}=-\Qhat_{\eta\zeta}.
\eqno{(3.5)}
$$
As well as in the case of $\commu{\Phat X}{\Phat Y}$ and $\symmp{\Phat X}{\Phat Y}$, then, one obtains the nonlocal representations of the projections of the commutator (2.18a) and the symmetrized product (2.18b) in the following way:
$$
\begin{array}{l}

\Phat\commu{X}{Y}\vs{12pt}\\

=\dps{\left.\sum^{\infty}_{n=0}C_{2n}(\Phat(\eta)\Phat(\zeta)(J^{\alpha\beta}\Zm_{\alpha}(\eta)\Zm_{\beta}(\zeta))^{2n}\commu{X(\eta)}{Y(\zeta)})\right|_{\eta=\zeta}}\vs{12pt}\\

\ts{24pt}-(-1)^s\dps{\left.\sum^{\infty}_{n=0}C_{2n+1}(\Phat(\eta)\Phat(\zeta)(J^{\alpha\beta}\Zm_{\alpha}(\eta)\Zm_{\beta}(\zeta))^{2n+1}2\symmp{X(\eta)}{Y(\zeta)})\right|_{\eta=\zeta}}\vs{12pt}\\

=\dps{\left(\Phat(\eta)\Phat(\zeta)\left.\sum^{\infty}_{n=0}(\f1{(2n)!}(\f{\hbar}{2i}\Qhat^t_{\eta\zeta})^{2n}\commu{X(\eta)}{Y(\zeta)}+\f1{(2n+1)!}(\f{\hbar}{2i}\Qhat^t_{\eta\zeta})^{2n+1}2\symmp{X(\eta)}{Y(\zeta)})\right)\right|_{\eta=\zeta}},

\end{array}
\eqno{(3.6a)}
$$
and similarly,
$$
\begin{array}{l}

\Phat\symmp{X}{Y}\vs{12pt}\\

=\dps{\left(\Phat(\eta)\Phat(\zeta)\left.\sum^{\infty}_{n=0}(\f1{(2n)!}(\f{\hbar}{2i}\Qhat^t_{\eta\zeta})^{2n}\symmp{X(\eta)}{Y(\zeta)}+\f1{(2n+1)!}(\f{\hbar}{2i}\Qhat^t_{\eta\zeta})^{2n+1}\f12\commu{X(\eta)}{Y(\zeta)})\right)\right|_{\eta=\zeta}}.

\end{array}
\eqno{(3.6b)}
$$


\subsection{Star-product representation of Projected operator Algebra}

\ts{12pt}The nonlocal representations (3.4a), (3.4b) and (3.6a), (3.6b) involve such nonlocal operator-product as $(\Qhat_{\eta\zeta})^nY(\zeta)X(\eta)$ through the nonlocal commutator $\commu{X(\eta)}{Y(\zeta)}$ and symmetrized product $\symmp{X(\eta)}{Y(\zeta)}$. From (3.5), $(\Qhat_{\eta\zeta})^nY(\zeta)X(\eta)$ satisfies 
$$
(\Qhat_{\eta\zeta})^nY(\zeta)X(\eta)=(-1)^n(\Qhat_{\eta\zeta})^nY(\eta)X(\zeta).
\eqno{(3.7)}
$$ 
With respect to the linear combinatios of nonlocal operator-products, then, one obtains the following formulas:
$$
(\Qhat_{\eta\zeta})^{2n}(X(\eta)Y(\zeta)\pm Y(\zeta)X(\eta))=(\Qhat_{\eta\zeta})^{2n}(X(\eta)Y(\zeta)\pm Y(\eta)X(\zeta)),
\eqno{(3.8a)}
$$

$$
(\Qhat_{\eta\zeta})^{2n+1}(X(\eta)Y(\zeta)\pm Y(\zeta)X(\eta))=(\Qhat_{\eta\zeta})^{2n+1}(X(\eta)Y(\zeta)\mp Y(\eta)X(\zeta)).
\eqno{(3.8b)}
$$

\subsubsection{Nonlocal representaion of $\commu{\Phat X}{\Phat Y}$ and $\symmp{\Phat X}{\Phat Y}$}

\ts{12pt}From the formulas (3.8a) and (3.8b), one obtains the followin formulas for the commutator and the symmetrized product of the projected operators $\Phat X$ and $\Phat Y$:
$$
\begin{array}{rcl}

\commu{\Phat X}{\Phat Y}
&=&
\dps{\Phat\left.\left(\sum^{\infty}_{n=0}\f1{n!}(\f{\hbar}{2i}\Qhat_{\eta\zeta})^n(X(\eta)Y(\zeta)-(-1)^{\varepsilon_X\varepsilon_Y}Y(\eta)X(\zeta))\right)\right|_{\eta=\zeta}}\vs{12pt}\\

&=&\dps{\Phat\left(\left.\exp(\f{\hbar}{2i}\Qhat_{\eta\zeta})X(\eta)Y(\zeta)\right|_{\eta=\zeta}-(-1)^{\varepsilon_X\varepsilon_Y}\left.\exp(\f{\hbar}{2i}\Qhat_{\eta\zeta})Y(\eta)X(\zeta)\right|_{\eta=\zeta}\right)},

\end{array}
\eqno{(3.9a)}
$$
$$
\begin{array}{rcl}

\symmp{\Phat X}{\Phat Y}
&=&
\dps{\Phat\left.\left(\sum^{\infty}_{n=0}\f1{n!}(\f{\hbar}{2i}\Qhat_{\eta\zeta})^n\f12(X(\eta)Y(\zeta)+(-1)^{\varepsilon_X\varepsilon_Y}Y(\eta)X(\zeta))\right)\right|_{\eta=\zeta}}\vs{12pt}\\

&=&
\dps{\Phat\left(\left.\f12(\exp(\f{\hbar}{2i}\Qhat_{\eta\zeta})X(\eta)Y(\zeta)\right|_{\eta=\zeta}+(-1)^{\varepsilon_X\varepsilon_Y}\left.\exp(\f{\hbar}{2i}\Qhat_{\eta\zeta})Y(\eta)X(\zeta)\right|_{\eta=\zeta})\right)}.
\end{array}
\eqno{(3.9b)}
$$ 

\subsubsection{Nonlocal representation of $\Phat \commu{X}{Y}$ and $\Phat\symmp{X}{Y}$}

From (3.5), $\Qhat^t_{\eta\zeta}$ satisfies the same formulas as (3.8a) and (3.8b). Therefore, one obtains the following formulas for the projections of $\commu{X}{Y}$ and $\symmp{X}{Y}$:
$$
\begin{array}{l}

\Phat \commu{X}{Y}
=
\dps{\left.\Phat(\eta)\Phat(\zeta)\sum^{\infty}_{n=0}\f1{n!}(\f{\hbar}{2i}\Qhat^t_{\eta\zeta})^n(X(\eta)Y(\zeta)-(-1)^{\varepsilon_X\varepsilon_Y}Y(\eta)X(\zeta))\right|_{\eta=\zeta}}\vs{12pt}\\

=\dps{\left.\Phat(\eta)\Phat(\zeta)\exp(\f{\hbar}{2i}\Qhat^t_{\eta\zeta})X(\eta)Y(\zeta)\right|_{\eta=\zeta}}
-\dps{(-1)^{\varepsilon_X\varepsilon_Y}\left.\Phat(\eta)\Phat(\zeta)\exp(\f{\hbar}{2i}\Qhat^t_{\eta\zeta})Y(\eta)X(\zeta)\right|_{\eta=\zeta}},

\end{array}
\eqno{(3.10a)}
$$
$$
\begin{array}{l}

\Phat \symmp{X}{Y}
=
\dps{\left.\left(\Phat(\eta)\Phat(\zeta)\sum^{\infty}_{n=0}\f1{n!}(\f{\hbar}{2i}\Qhat^t_{\eta\zeta})^n\f12(X(\eta)Y(\zeta)+(-1)^{\varepsilon_X\varepsilon_Y}Y(\eta)X(\zeta))\right)\right|_{\eta=\zeta}}\vs{12pt}\\

=\dps{\left.\f12\left(\Phat(\eta)\Phat(\zeta)\exp(\f{\hbar}{2i}\Qhat^t_{\eta\zeta})X(\eta)Y(\zeta)\right|_{\eta=\zeta}\right.}+\dps{\left.(-1)^{\varepsilon_X\varepsilon_Y}\left.\Phat(\eta)\Phat(\zeta)\exp(\f{\hbar}{2i}\Qhat^t_{\eta\zeta})Y(\eta)X(\zeta)\right|_{\eta=\zeta}\right)}.

\end{array}
\eqno{(3.10b)}
$$
 
\subsection{Star repesentation of Projected operator Algebra}

\ts{12pt}The structures of the commutator and  symmetrized product formulas (3.9a)-(3.10b) suggest that these formulas can be naturally reformulated in terms of the so-called star-product like the Moyal star-product. \\
\ts{12pt}For this purpose, we first introduce new operator-products in the quantum constraint systems as follows:
$$
X\star Y=\left.\exp(\f{\hbar}{2i}\Qhat_{\eta\zeta})X(\eta)Y(\zeta)\right|_{\eta=\zeta},
\eqno{(3.11)}
$$  
which we shall call the constraint $\star$ -product, and
$$
X\pscrp\star   Y=\left.\left(\Phat(\eta)\Phat(\zeta)\exp(\f{\hbar}{2i}\Qhat^t_{\eta\zeta})X(\eta)Y(\zeta)\right)\right|_{\eta=\zeta},
\eqno{(3.12)}
$$
which, the constraint $\pscrp\star$ -product.\\
\ts{12pt}We next define the commutators and the symmetrized products by these star-products as follows:
$$
\begin{array}{ll}
\commu{X}{Y}_{\star}=X\star Y-(-1)^{\varepsilon_X\varepsilon_Y}Y\star X&(\star \mbox{-commutator}),\vs{12pt}\\
\symmp{X}{Y}_{\star}=\dps{\f12}(X\star Y+(-1)^{\varepsilon_X\varepsilon_Y}Y\star  X)&(\star \mbox{-symmetrized product})
\end{array}
\eqno{(3.13)} 
$$
and
$$
\begin{array}{ll}
\commu{X}{Y}_{\pscrp\star}=X\pscrp\star Y-(-1)^{\varepsilon_X\varepsilon_Y}Y\pscrp\star X&(\pscrp\star \mbox{-commutator}),\vs{12pt}\\
\symmp{X}{Y}_{\pscrp\star}=\dps{\f12}(X\pscrp\star   Y+(-1)^{\varepsilon_X\varepsilon_Y}Y\pscrp\star X)&(\pscrp\star  \mbox{-symmetrized product}).
\end{array}
\eqno{(3.14)}
$$ 
Then, it is obvious that the commutators in (3.13) and (3.14) satisfy the graded Jacobi identities (see Appendix B).
\\

Thus, the nonlocal representations (3.9a)-(3.10b) are reformulated by the star-product formulas (3.13) and(3.14)  in the following way:
$$
\begin{array}{rcl}

\commu{\Phat X}{\Phat Y}
&=&
\dps{\Phat\left(\left.\exp(\f{\hbar}{2i}\Qhat_{\eta\zeta})X(\eta)Y(\zeta)\right|_{\eta=\zeta}-(-1)^{\varepsilon_X\varepsilon_Y}\left.\exp(\f{\hbar}{2i}\Qhat_{\eta\zeta})Y(\eta)X(\zeta)\right|_{\eta=\zeta}\right)}\vs{12pt}\\
&=&
\Phat(X\star Y-(-1)^{\varepsilon_X\varepsilon_Y}Y\star X)=\Phat\commu{X}{Y}_{\star},
\end{array}
\eqno{(3.15a)}
$$

$$
\begin{array}{rcl}
\symmp{\Phat X}{\Phat Y}
&=&
\dps{\Phat\left(\left.\f12(\exp(\f{\hbar}{2i}\Qhat_{\eta\zeta})X(\eta)Y(\zeta)\right|_{\eta=\zeta}+(-1)^{\varepsilon_X\varepsilon_Y}\left.\exp(\f{\hbar}{2i}\Qhat_{\eta\zeta})Y(\eta)X(\zeta)\right|_{\eta=\zeta})\right)}\vs{12pt}\\
&=&
\Phat\f12(X\star Y+(-1)^{\varepsilon_X\varepsilon_Y}Y\star X)=\Phat\symmp{X}{Y}_{\star},
\end{array}
\eqno{(3.15b)}
$$  
and
$$
\begin{array}{rcl}
\Phat\commu{X}{Y}
&=&
\dps{\left.\Phat(\eta)\Phat(\zeta)\exp(\f{\hbar}{2i}\Qhat^t_{\eta\zeta})X(\eta)Y(\zeta)\right|_{\eta=\zeta}}
-\dps{(-1)^{\varepsilon_X\varepsilon_Y}\left.\Phat(\eta)\Phat(\zeta)\exp(\f{\hbar}{2i}\Qhat^t_{\eta\zeta})Y(\eta)X(\zeta)\right|_{\eta=\zeta}}\vs{12pt}\\
&=&
X\pscrp\star Y-(-1)^{\varepsilon_X\varepsilon_Y}Y\pscrp\star X=\commu{X}{Y}_{\pscrp\star},
\end{array}
\eqno{(3.16a)}
$$

$$
\begin{array}{lcl}
\ts{-24pt}\Phat\symmp{X}{Y}
&=&
\dps{\left.\f12\left(\Phat(\eta)\Phat(\zeta)\exp(\f{\hbar}{2i}\Qhat^t_{\eta\zeta})X(\eta)Y(\zeta)\right|_{\eta=\zeta}\right.}+\dps{\left.(-1)^{\varepsilon_X\varepsilon_Y}\left.\Phat(\eta)\Phat(\zeta)\exp(\f{\hbar}{2i}\Qhat^t_{\eta\zeta})Y(\eta)X(\zeta)\right|_{\eta=\zeta}\right)}\vs{12pt}\\
&=&
\dps{\f12}(X\pscrp\star Y+(-1)^{\varepsilon_X\varepsilon_Y}Y\pscrp\star X)=\symmp{X}{Y}_{\pscrp\star}.
\end{array}
\eqno{(3.16b)}
$$
\ts{12pt}Successively applying the formulas (3.15a)-(3.16b), the commutator of the projected operators $\Phat X$ and $\Phat Y$ can be represented in the form of the power series of $\hbar$ as follows:
$$
\commu{\Phat X}{\Phat Y}=i\hbar\sum^{\infty}_{n=0}\hbar^{2n}C_n(\calC^*),
\eqno{(3.17)}
$$
where the first term is the quantized form of the Dirac bracket and the rest higher order terms of $\hbar$ are the quantum corrections caused by the noncommutativity among the ACCS and the operators $X,Y$. Similarly, the projection of the symmetrized product, $\Phat\symmp{X}{Y}$ is represented in the following form:
$$
\Phat\symmp{X}{Y}= \sum^{\infty}_{n=0}\hbar^{2n}S_n(\calC^*),
\eqno{(3.18)}
$$
which contains the qunatum corrections in the form of the power series of $\hbar^{2n}$ $(n\geq 1)$.

\section{Discussion and concluding remarks}
\ts{12pt}We have proposed the new types of star-products in the quantization of the second-class constraint systems with the operator formalism. Although the ordinary star-product
$$
f(z)\star g(z)=\left.\exp(\f{i\hbar}2\theta^{\mu\nu}\f{\partial}{\partial z_1^{\mu}}\f{\partial}{\partial z_2^{\nu}})f(z_1^{\mu})g(z_2^{\nu})\right|_{z_1=z_2=z}
\eqno{(4.1)}
$$
is expressed in terms of the original CCS,  $\{z^{\mu}=(q^i,p_i:i=1,\cdots,N)\}$, with the nonlocal representation, the star-products (3.11) and (3.12) are described in terms of the {\it hyper}-operators $\Zm_{\alpha}$ $(\alpha=1,\cdots,2M)$ of the ACCS, which is the subset the modified CCS,  $\calC^*\oplus\{Z_\alpha\}$.
Then, one sees that the new star-products defined by (3.11) and (3.12) are the projections of the ordinary star-products to $\calC^*$ in the operator formalism.\\
\ts{12pt}We have thus shown that the commutator formulas and the symmetrized product ones in the POM are naturally reformulated by these star-products:\\
For $X^*,\ Y^*\ \in \calC^*$, 
$$
\begin{array}{l}
\commu{X^*}{Y^*}=\commu{\Phat X}{\Phat Y}=\Phat\commu{X}{Y}_{\star},\vs{12pt}\\
\symmp{X^*}{Y^*}=\symmp{\Phat X}{\Phat Y}=\Phat\symmp{X}{Y}_{\star},
\end{array}
\eqno{(4.2)}
$$
and
$$
\begin{array}{l}
\commu{X}{Y}^*=\Phat\commu{X}{Y}=\commu{X}{Y}_{\pscrp\star},\vs{12pt}\\
\symmp{X}{Y}^*=\Phat\symmp{X}{Y}=\symmp{X}{Y}_{\pscrp\star}.
\end{array}
\eqno{(4.3)}
$$

Successively applying the these formulas, it has been shown that the
commutators among the projected operators and the 
operator products contain the quantum effect caused by the noncommutativity among the ACCS and the operators in the form of the power series of $\hbar$. 

\newpage

\appendix

\section{Some Projection Formulas in POM}

\ts{12pt}In order to calculate the projections of the commutators and the symmetrized products, which contain such a type of $X(\Zp)^nY$, some availabe formulas are presented. \\
\ts{12pt}From (2.10) and (2.14), the following formulas are obtained by the straightforward calculations:
$$
\begin{array}{ll}
\dps{\Phat[X,\ \Zp_{\alpha_1}\cdots\Zp_{\alpha_{2n}}Y]=(-1)^{ns}\left(\f{\hbar}{2i}\right)^{2n}\Phat[\Zm_{\alpha_{2n}}\cdots\Zm_{\alpha_1}X, \ Y]}\hs{90pt}&(\mbox{A}1)
\vs{24pt}\\
\dps{\Phat[X,\ \Zp_{\alpha_1}\cdots\Zp_{\alpha_{2n+1}}Y]=2(-1)^{\varepsilon_Xs+ns}\left(\f{\hbar}{2i}\right)^{2n+1}\Phat\{\Zm_{\alpha_{2n+1}}\cdots\Zm_{\alpha_1}X, \ Y\}}&
(\mbox{A}2)\vs{24pt}\\
\dps{\Phat\{X,\ \Zp_{\alpha_1}\cdots\Zp_{\alpha_{2n}}Y\}=(-1)^{ns}\left(\f{\hbar}{2i}\right)^{2n}\Phat\{\Zm_{\alpha_{2n}}\cdots\Zm_{\alpha_1}X, \ Y\}}&(\mbox{A}3)
\vs{24pt}\\
\dps{\Phat\{X,\ \Zp_{\alpha_1}\cdots\Zp_{\alpha_{2n+1}}Y\}=\f12(-1)^{\varepsilon_Xs+ns}\left(\f{\hbar}{2i}\right)^{2n+1}\Phat[\Zm_{\alpha_{2n+1}}\cdots\Zm_{\alpha_1}X, \ Y]}&
(\mbox{A}4)
\end{array}
$$

\section{Graded Jacobi identity }

\ts{12pt}For any operators $A$, $B$ and $C$, the commutator $\commu{A}{\commu{B}{C}_{\star}}_{\star}$ is defined by
$$
\begin{array}{l}
\commu{A}{\commu{B}{C}_{\star}}_{\star}=A\star(B\star C)-(-1)^{\varepsilon_B\varepsilon_C}A\star(C\star B)\vs{12pt}\\
-(-1)^{\varepsilon_A\varepsilon_B+\varepsilon_C\varepsilon_A}(B\star C)\star A+(-1)^{\varepsilon_A\varepsilon_B+\varepsilon_B\varepsilon_C+\varepsilon_C\varepsilon_A}(C\star B)\star A\vs{12pt}\\
\hs{180pt}(\mbox{cyclic for }A,B,C),
\end{array}
\eqno{(\mbox{B}1)} 
$$
where
$$
\begin{array}{rcl}
A\star(B\star C)&=&\dps{
\left.\exp(\f{\hbar}{2i}\Qhat_{\eta\zeta})A(\eta)(B\star C)(\zeta)\right|_{\eta=\zeta}
}\vs{12pt}\\
&=&\dps{
\left.\exp(\f{\hbar}{2i}\Qhat_{\eta\zeta})A(\eta)\left(\left.\exp(\f{\hbar}{2i}\Qhat_{\eta'\zeta'})B(\eta')C(\zeta')\right|_{\eta'=\zeta'=\zeta}\right)\right|_{\eta=\zeta}
},
\end{array}
\eqno{(\mbox{B2})}
$$
and 
$$
\begin{array}{rcl}
(A\star B)\star C&=&\dps{
\left.\exp(\f{\hbar}{2i}\Qhat_{\eta\zeta})(A\star B)(\eta)\star C(\zeta)\right|_{\eta=\zeta}
}\vs{12pt}\\
&=&\dps{
\left.\exp(\f{\hbar}{2i}\Qhat_{\eta\zeta})\left(\left.\exp(\f{\hbar}{2i}\Qhat_{\eta'\zeta'})A(\eta')B(\zeta')\right|_{\eta'=\zeta'=\eta}\right)C(\zeta)\right|_{\eta=\zeta}
}.
\end{array}
\eqno{(\mbox{B3})}
$$
From (B1), then, the $\star$-commutator satisfies the graded Jacobi identy
$$
(-1)^{\varepsilon_C\varepsilon_A}\commu{A}{\commu{B}{C}_{\star}}_{\star}+(-1)^{\varepsilon_A\varepsilon_B}\commu{B}{\commu{C}{A}_{\star}}_{\star}+(-1)^{\varepsilon_B\varepsilon_C}\commu{C}{\commu{A}{B}_{\star}}_{\star}=0.
\eqno{(\mbox{B}4)}
$$
\\
\ts{12pt}The commutator $\commu{A}{\commu{B}{C}_{\pscrp\star}}_{\pscrp\star}$ takes the same structure to the $\star$-commutator:
$$
\begin{array}{l}
\commu{A}{\commu{B}{C}_{\pscrp\star}}_{\pscrp\star}=A\pscrp\star(B\pscrp\star C)-(-1)^{\varepsilon_B\varepsilon_C}A\pscrp\star(C\pscrp\star B)\vs{12pt}\\
-(-1)^{\varepsilon_A\varepsilon_B+\varepsilon_C\varepsilon_A}(B\pscrp\star C)\pscrp\star A+(-1)^{\varepsilon_A\varepsilon_B+\varepsilon_B\varepsilon_C+\varepsilon_C\varepsilon_A}(C\pscrp\star B)\pscrp\star A\vs{12pt}\\
\hs{180pt}(\mbox{cyclic for }A,B,C)
\end{array}
\eqno{(\mbox{B}5)} 
$$
with
$$
\begin{array}{l}
A\pscrp\star(B\pscrp\star C)=\dps{\Phat(\eta)\Phat(\zeta)
\left.\exp(\f{\hbar}{2i}\Qhat_{\eta\zeta})A(\eta)(B\pscrp\star C)(\zeta)\right|_{\eta=\zeta}
}\vs{12pt}\\
=\dps{
\left.\Phat(\eta)\Phat(\zeta)\exp(\f{\hbar}{2i}\Qhat_{\eta\zeta})A(\eta)\left(\left.\Phat(\eta')\Phat(\zeta')\exp(\f{\hbar}{2i}\Qhat_{\eta'\zeta'})B(\eta')C(\zeta')\right|_{\eta'=\zeta'=\zeta}\right)\right|_{\eta=\zeta}
}.
\end{array}
\eqno{(\mbox{B}6)}
$$
ThenAthe $\pscrp\star$-commutator also satisfies the the graded Jacobi identy
$$
(-1)^{\varepsilon_C\varepsilon_A}\commu{A}{\commu{B}{C}_{\pscrp\star}}_{\pscrp\star}+(-1)^{\varepsilon_A\varepsilon_B}\commu{B}{\commu{C}{A}_{\pscrp\star}}_{\pscrp\star}+(-1)^{\varepsilon_B\varepsilon_C}\commu{C}{\commu{A}{B}_{\pscrp\star}}_{\pscrp\star}=0.
\eqno{(\mbox{B}7)}
$$

\end{document}